\RequirePackage{lineno}
\documentclass[prd,twocolumn,showpacs,superscriptaddress,floatfix,linenumbers,nofootinbib,fleqn]{revtex4}
\usepackage{graphicx}  
\usepackage{dcolumn}   
\usepackage{bm}        
\usepackage{amssymb}   
\usepackage{amsmath}
\usepackage{epsfig}
\usepackage{subfigure}
\usepackage{psfrag}
\usepackage{xspace}
\usepackage{multirow}
\usepackage{color}
\usepackage{lineno}
\usepackage{mathrsfs}
\usepackage{enumerate}

\newcommand{\ttbar}     {\mbox{\ensuremath{t\bar{t}}}}
\newcommand{\ppbar}     {\mbox{\ensuremath{p\bar{p}}}}
\newcommand{\qqbar}     {\mbox{\ensuremath{q\bar{q}}}}

\newcommand{\pt}        {\mbox{$p_T$}}
\newcommand{\met}       {\mbox{\ensuremath{\slash\kern-.55emp_{T}}}}
\newcommand{\metsig}    {\mbox{\ensuremath{\sigma_{\slash\kern-.4emp_{T}}}}}
\newcommand{\ljets}     {\mbox{$\ell$+jets}}
\newcommand{\dilepton}  {\mbox{$\ell\ell$}}
\newcommand{\herwig}    {{\sc herwig}}
\newcommand{\pythia}    {{\sc pythia}}
\newcommand{\alpgen}    {{\sc alpgen}}
\newcommand{\geant}     {{\sc geant3}}
\newcommand{\mcatnlo}   {{\sc mc@nlo}}
\newcommand{\vecbos}{{\sc vecbos}\xspace}

\newcommand{\Z}         {\mbox{$Z/\gamma^\star$}}
\newcommand{\zlljets}{\ensuremath{(\Z\rightarrow\ell\ell)+\rm jets}\xspace}
\newcommand{\etadet} 	{\ensuremath{\eta_{\text{det}}}}

\newcommand{\kjes}      {\ensuremath{k_{\rm JES}}}

\newcommand{\zjets}{\ensuremath{\Z+\rm jets}\xspace}

\newcommand{\mtop}{\ensuremath{m_{ t}}\xspace}

\newcommand{\MeV}{\ensuremath{\mathrm{Me\kern-0.1em V}}\xspace}
\newcommand{\GeV}{\ensuremath{\mathrm{Ge\kern-0.1em V}}\xspace}
\begin{document}



\newcommand{\lumi}      {9.7~fb$^{-1}$}
\newcommand{\dzero}{D\O\xspace}

\newcommand{\mmeas}{\ensuremath{173.93}}
\newcommand{\mstat}{\ensuremath{1.61}}
\newcommand{\msyst}{\ensuremath{0.88}}
\newcommand{\merr}{\ensuremath{1.84}}
\newcommand{\result}{\ensuremath{m_t = mmeas \pm mstat {\rm (stat)} \pm msyst {\rm (syst)}~\GeV}}

\mbox{FERMILAB-PUB-16-219-E \hspace{20mm}{\em Published in Phys. Rev. D as DOI: 10.1103/PhysRevD.94.032004}}

\title{Measurement of the Top Quark Mass Using the Matrix Element Technique in Dilepton Final States.}
\vspace*{0.1cm}

%
\affiliation{LAFEX, Centro Brasileiro de Pesquisas F\'{i}sicas, Rio de Janeiro, RJ 22290, Brazil}
\affiliation{Universidade do Estado do Rio de Janeiro, Rio de Janeiro, RJ 20550, Brazil}
\affiliation{Universidade Federal do ABC, Santo Andr\'e, SP 09210, Brazil}
\affiliation{University of Science and Technology of China, Hefei 230026, People's Republic of China}
\affiliation{Universidad de los Andes, Bogot\'a, 111711, Colombia}
\affiliation{Charles University, Faculty of Mathematics and Physics, Center for Particle Physics, 116 36 Prague 1, Czech Republic}
\affiliation{Czech Technical University in Prague, 116 36 Prague 6, Czech Republic}
\affiliation{Institute of Physics, Academy of Sciences of the Czech Republic, 182 21 Prague, Czech Republic}
\affiliation{Universidad San Francisco de Quito, Quito, Ecuador}
\affiliation{LPC, Universit\'e Blaise Pascal, CNRS/IN2P3, Clermont, F-63178 Aubi\`ere Cedex, France}
\affiliation{LPSC, Universit\'e Joseph Fourier Grenoble 1, CNRS/IN2P3, Institut National Polytechnique de Grenoble, F-38026 Grenoble Cedex, France}
\affiliation{CPPM, Aix-Marseille Universit\'e, CNRS/IN2P3, F-13288 Marseille Cedex 09, France}
\affiliation{LAL, Univ. Paris-Sud, CNRS/IN2P3, Universit\'e Paris-Saclay, F-91898 Orsay Cedex, France}
\affiliation{LPNHE, Universit\'es Paris VI and VII, CNRS/IN2P3, F-75005 Paris, France}
\affiliation{CEA Saclay, Irfu, SPP, F-91191 Gif-Sur-Yvette Cedex, France}
\affiliation{IPHC, Universit\'e de Strasbourg, CNRS/IN2P3, F-67037 Strasbourg, France}
\affiliation{IPNL, Universit\'e Lyon 1, CNRS/IN2P3, F-69622 Villeurbanne Cedex, France and Universit\'e de Lyon, F-69361 Lyon CEDEX 07, France}
\affiliation{III. Physikalisches Institut A, RWTH Aachen University, 52056 Aachen, Germany}
\affiliation{Physikalisches Institut, Universit\"at Freiburg, 79085 Freiburg, Germany}
\affiliation{II. Physikalisches Institut, Georg-August-Universit\"at G\"ottingen, 37073 G\"ottingen, Germany}
\affiliation{Institut f\"ur Physik, Universit\"at Mainz, 55099 Mainz, Germany}
\affiliation{Ludwig-Maximilians-Universit\"at M\"unchen, 80539 M\"unchen, Germany}
\affiliation{Panjab University, Chandigarh 160014, India}
\affiliation{Delhi University, Delhi-110 007, India}
\affiliation{Tata Institute of Fundamental Research, Mumbai-400 005, India}
\affiliation{University College Dublin, Dublin 4, Ireland}
\affiliation{Korea Detector Laboratory, Korea University, Seoul, 02841, Korea}
\affiliation{CINVESTAV, Mexico City 07360, Mexico}
\affiliation{Nikhef, Science Park, 1098 XG Amsterdam, the Netherlands}
\affiliation{Radboud University Nijmegen, 6525 AJ Nijmegen, the Netherlands}
\affiliation{Joint Institute for Nuclear Research, Dubna 141980, Russia}
\affiliation{Institute for Theoretical and Experimental Physics, Moscow 117259, Russia}
\affiliation{Moscow State University, Moscow 119991, Russia}
\affiliation{Institute for High Energy Physics, Protvino, Moscow region 142281, Russia}
\affiliation{Petersburg Nuclear Physics Institute, St. Petersburg 188300, Russia}
\affiliation{Instituci\'{o} Catalana de Recerca i Estudis Avan\c{c}ats (ICREA) and Institut de F\'{i}sica d'Altes Energies (IFAE), 08193 Bellaterra (Barcelona), Spain}
\affiliation{Uppsala University, 751 05 Uppsala, Sweden}
\affiliation{Taras Shevchenko National University of Kyiv, Kiev, 01601, Ukaine}
\affiliation{Lancaster University, Lancaster LA1 4YB, United Kingdom}
\affiliation{Imperial College London, London SW7 2AZ, United Kingdom}
\affiliation{The University of Manchester, Manchester M13 9PL, United Kingdom}
\affiliation{University of Arizona, Tucson, Arizona 85721, USA}
\affiliation{University of California Riverside, Riverside, California 92521, USA}
\affiliation{Florida State University, Tallahassee, Florida 32306, USA}
\affiliation{Fermi National Accelerator Laboratory, Batavia, Illinois 60510, USA}
\affiliation{University of Illinois at Chicago, Chicago, Illinois 60607, USA}
\affiliation{Northern Illinois University, DeKalb, Illinois 60115, USA}
\affiliation{Northwestern University, Evanston, Illinois 60208, USA}
\affiliation{Indiana University, Bloomington, Indiana 47405, USA}
\affiliation{Purdue University Calumet, Hammond, Indiana 46323, USA}
\affiliation{University of Notre Dame, Notre Dame, Indiana 46556, USA}
\affiliation{Iowa State University, Ames, Iowa 50011, USA}
\affiliation{University of Kansas, Lawrence, Kansas 66045, USA}
\affiliation{Louisiana Tech University, Ruston, Louisiana 71272, USA}
\affiliation{Northeastern University, Boston, Massachusetts 02115, USA}
\affiliation{University of Michigan, Ann Arbor, Michigan 48109, USA}
\affiliation{Michigan State University, East Lansing, Michigan 48824, USA}
\affiliation{University of Mississippi, University, Mississippi 38677, USA}
\affiliation{University of Nebraska, Lincoln, Nebraska 68588, USA}
\affiliation{Rutgers University, Piscataway, New Jersey 08855, USA}
\affiliation{Princeton University, Princeton, New Jersey 08544, USA}
\affiliation{State University of New York, Buffalo, New York 14260, USA}
\affiliation{University of Rochester, Rochester, New York 14627, USA}
\affiliation{State University of New York, Stony Brook, New York 11794, USA}
\affiliation{Brookhaven National Laboratory, Upton, New York 11973, USA}
\affiliation{Langston University, Langston, Oklahoma 73050, USA}
\affiliation{University of Oklahoma, Norman, Oklahoma 73019, USA}
\affiliation{Oklahoma State University, Stillwater, Oklahoma 74078, USA}
\affiliation{Oregon State University, Corvallis, Oregon 97331, USA}
\affiliation{Brown University, Providence, Rhode Island 02912, USA}
\affiliation{University of Texas, Arlington, Texas 76019, USA}
\affiliation{Southern Methodist University, Dallas, Texas 75275, USA}
\affiliation{Rice University, Houston, Texas 77005, USA}
\affiliation{University of Virginia, Charlottesville, Virginia 22904, USA}
\affiliation{University of Washington, Seattle, Washington 98195, USA}
\author{V.M.~Abazov} \affiliation{Joint Institute for Nuclear Research, Dubna 141980, Russia}
\author{B.~Abbott} \affiliation{University of Oklahoma, Norman, Oklahoma 73019, USA}
\author{B.S.~Acharya} \affiliation{Tata Institute of Fundamental Research, Mumbai-400 005, India}
\author{M.~Adams} \affiliation{University of Illinois at Chicago, Chicago, Illinois 60607, USA}
\author{T.~Adams} \affiliation{Florida State University, Tallahassee, Florida 32306, USA}
\author{J.P.~Agnew} \affiliation{The University of Manchester, Manchester M13 9PL, United Kingdom}
\author{G.D.~Alexeev} \affiliation{Joint Institute for Nuclear Research, Dubna 141980, Russia}
\author{G.~Alkhazov} \affiliation{Petersburg Nuclear Physics Institute, St. Petersburg 188300, Russia}
\author{A.~Alton$^{a}$} \affiliation{University of Michigan, Ann Arbor, Michigan 48109, USA}
\author{A.~Askew} \affiliation{Florida State University, Tallahassee, Florida 32306, USA}
% ...
%
%
\collaboration{The D0 Collaboration\footnote{With visitors from
$^{a}$Augustana College, Sioux Falls, SD 57197, USA,
$^{b}$The University of Liverpool, Liverpool L69 3BX, UK,
$^{c}$Deutshes Elektronen-Synchrotron (DESY), Notkestrasse 85, Germany,
$^{d}$CONACyT, M-03940 Mexico City, Mexico,
$^{e}$SLAC, Menlo Park, CA 94025, USA,
$^{f}$University College London, London WC1E 6BT, UK,
$^{g}$Centro de Investigacion en Computacion - IPN, CP 07738 Mexico City, Mexico,
$^{h}$Universidade Estadual Paulista, S\~ao Paulo, SP 01140, Brazil,
$^{i}$Karlsruher Institut f\"ur Technologie (KIT) - Steinbuch Centre for Computing (SCC),
D-76128 Karlsruhe, Germany,
$^{j}$Office of Science, U.S. Department of Energy, Washington, D.C. 20585, USA,
$^{k}$American Association for the Advancement of Science, Washington, D.C. 20005, USA,
$^{l}$Kiev Institute for Nuclear Research (KINR), Kyiv 03680, Ukraine,
$^{m}$University of Maryland, College Park, MD 20742, USA,
$^{n}$European Orgnaization for Nuclear Research (CERN), CH-1211 Geneva, Switzerland
and
$^{o}$Purdue University, West Lafayette, IN 47907, USA.
$^{\ddag}$Deceased.
}} \noaffiliation
\vskip 0.25cm

\date{August 18, 2016}
\begin{abstract}
We present a measurement of the top quark mass in \ppbar\ collisions at  a center-of-mass energy  of 1.96~TeV at the Fermilab Tevatron collider.
The data were collected by the D0 experiment corresponding to an integrated luminosity of \lumi.
The matrix element technique is applied to \ttbar\ events in the final state containing leptons (electrons or muons) with high transverse momenta 
and at least two jets.
The calibration of the jet energy scale determined in the lepton + jets
final state of \ttbar\ decays is applied to jet energies. This correction
provides a substantial reduction in systematic uncertainties. 
We obtain a top quark mass of  $m_t = \mmeas \pm \merr$~GeV.
  
\end{abstract}


\maketitle 

\section{Introduction}


\label{sec:intro}

The top quark is the heaviest elementary particle of the standard model (SM)~\cite{Abazov:2014dpa,Abazov:2015spa,Tevatron:2014cka,CMSmt:2016,ATLAS:2014wva}.
Its mass ($m_t$) is a free parameter of the SM Lagrangian that is not predicted from first principles. 
The top quark was discovered in 1995  by the CDF and D0 Collaborations
at the Tevatron $p\bar{p}$ collider at Fermilab~\cite{Abachi:1995iq,Abe:1995hr}.
Despite the fact that the top quark decays weakly, its large mass leads to a  very short lifetime
of approximately $ 5\cdot10^{-25}$~s~\cite{Jezabek:1987nf, Jezabek:1988iv,Abazov2012n}.
It decays into a $W$~boson and a $b$ quark before hadronizing, 
a process that has a characteristic time scale of
$1/\Lambda_{\rm{QCD}} \approx (200~{\rm MeV})^{-1}$,
equivalent to $\tau_{\rm{had}} \approx 3.3\cdot10^{-24}$~s,
where $\Lambda_{\rm{QCD}}$ is the fundamental scale of
quantum chromodynamics (QCD).  
This provides an opportunity to measure the mass of the top quark with high precision due to possibility of reconstructing the top quark parameters using its decay particles. 

At the Tevatron,
top quarks are produced mainly as \ttbar\  pairs through the strong interaction.
At leading order (LO) in perturbative QCD, a pair of top quarks is produced 
via quark-antiquark (\qqbar) annihilation with a probability of about 85\%~\cite{Bernreuther:2004jv,Czakon:2013tha}, or via gluon-gluon ($gg$) fusion. 

Final states of \ttbar\  production  are classified according to the decays of the two $W$~bosons. This results in final states with two, one, or no leptons, which are referred to as the dilepton (\dilepton), lepton + jets (\ljets), and all-jet channels, respectively.
In this measurement we use events in the dilepton final state where both $W$~bosons decay to leptons:
$ t\bar{t} \to  W^+b\ W^-\bar{b} \to \ell^+\nu_\ell b\ \ell^- \bar{\nu_\ell} \bar{b}$. 
More specifically, we consider three combinations of leptons, $ee$, $e\mu$, and $\mu\mu$, including also 
electrons and muons from leptonic decays of $\tau$ leptons, $W \to \tau\nu_\tau \to \ell \nu_\ell \nu_\tau$.
   We present an updated measurement of the top quark mass in the dilepton channel using the matrix
element (ME) approach ~\cite{Abazov:2004cs}.
This measurement improves the previous result using the matrix element technique with 5.3~fb$^{-1}$ of integrated luminosity~\cite{Abazov2011c} by a factor of 1.6, where the statistical uncertainty is improved by a factor of 1.1 and systematic uncertainty by a factor of 2.7.
The most precise \mtop measurement by D0 experiment based on this method was performed in \ljets\ analysis ~\cite{Abazov:2014dpa,Abazov:2015spa}. The CMS Collaboration has applied a different approach for measuring $m_t$ in the dilepton channel, obtaining a precision of 1.23 GeV~\cite{CMSmt:2016}.

This measurement uses the entire data set accumulated by the D0 experiment during Run II of the Fermilab 
Tevatron collider, corresponding to an integrated luminosity of~\lumi. 
We use  the final D0 jet energy scale (JES) corrections and the refined corrections of the
$b$ quark jet energy scale~\cite{Abazov:2013hda}. The measurement is performed with a blinded approach, as described in Section \ref{sec:app2data}.
Similarly to the recent top mass measurement in the dilepton final state using a
neutrino weighting technique~\cite{D0:2015dxa}, we correct jet energies by a calibration factor
obtained in the top quark mass measurement in the \ljets\ analysis~\cite{Abazov:2014dpa,Abazov:2015spa}.

\section{Detector and event samples}
\label{sec:detector}
\subsection{D0 detector}
The D0 detector is described in detail in
Refs.~\cite{Abazov2006l,Abazov:2005uk,Abolins2008,Angstadt2010,Ahmed:2010fx,Casey:2012rr,Bezzubov:2014jka}. 
It has a  central tracking system consisting of a
silicon microstrip tracker and a central fiber tracker,
both located within a 2~T superconducting solenoidal
magnet. The central tracking system is designed to optimize tracking and
vertexing at detector pseudorapidities  of 
$|\etadet|<2.5$.\footnote{The pseudorapidity is defined as $\eta = - \ln [\tan(\theta/2)]$, where $\theta$ is the polar angle of the reconstructed particle originating from a primary vertex relative to the proton beam direction. Detector pseudorapidity \etadet\ is defined relative to center of the detector instead of the primary vertex.}
A liquid-argon sampling calorimeter has a
central section (CC) covering $|\etadet|$ up to
$\approx 1.1$, and two end calorimeters (EC)  that extend coverage
to $|\etadet|\approx 4.2$, with all three housed in separate
cryostats. An outer muon system, with pseudorapidity coverage of $|\etadet|<2$,
consists of a layer of tracking detectors and scintillation trigger
counters in front of 1.8~T iron toroids, followed by two similar layers
after the toroids.

The sample of $p\bar p$ collision data considered in this analysis is split into 
four data-taking periods: ``Run~IIa'', ``Run~IIb1'', ``Run~IIb2'', and 
``Run~IIb3'' with the corresponding integrated luminosities given in Table I. 
All event simulations are split according to these epochs to better model changes of detector response with time, such as the addition of an additional SMT layer~\cite{Angstadt2010}or the reconstruction algorithm performance variations due to increasing luminosity~\cite{Abazov:2013xpp}.

\subsection{Object identification}
Top pair events in the dilepton channel contain two isolated charged leptons, 
two $b$ quark jets,  and a significant imbalance in transverse momentum (\met) due to escaping neutrinos.

Electrons are identified as energy clusters in the calorimeter within a cone of radius ${\cal R}=\sqrt{(\Delta\eta)^2+(\Delta\phi)^2}=0.2$ 
(where $\phi$ is the azimuthal angle) that are
consistent in their longitudinal and transverse profiles
with expectations from electromagnetic showers.
More than 90\% of the energy of an electron candidate must be deposited  in the 
electromagnetic part of the calorimeter.
The electron is required to be isolated by demanding that less than $20$\% of its energy is deposited in an
annulus of $0.2 < {\cal R} < 0.4$ around its direction.
This cluster has to be matched to a track reconstructed in the central tracking system.
We consider electrons in the CC with  $|\etadet| <1.1$ and in the EC with $1.5 < |\etadet| < 2.5$.
The transverse momenta of electrons ($p_T^e$) must be greater than $15$ GeV.
In addition, we use a multivariate discriminant based on tracking 
and calorimeter information to reject jets misidentified as electrons. 
It has an electron selection efficiency between 75\% and 80\%, depending on the data taking period, rapidity of the electron, and number of jets in the event. The rejection rate for jets is approximately
 $ 96$\%.

Muons are identified~\cite{Abazov:2013xpp} as segments in at least one layer 
of the  muon system that are
matched to tracks reconstructed in the central tracking system. 
Reconstructed muons must have $p_T>15$~GeV, $|\eta|<2$, and satisfy the two following isolation criteria.
First, the transverse energy deposited in the calorimeter annulus $0.1 < {\cal R} < 0.4$
around the muon  ($E_T^{\mu,\text{iso}}$)
must be less than 15\% of the transverse momentum of the muon ($p^{\mu}_T$).
Secondly, the sum of the transverse momenta of the tracks in 
a cone of radius ${\cal R}=0.5$ around the muon track 
in the central tracking system ($p_T^{\mu,\text{iso}}$) must be less than 15\% of $p^{\mu}_T$.

Jets are identified as energy clusters in the electromagnetic and hadronic parts of the calorimeter,
reconstructed using an iterative mid-point cone algorithm with radius ${\cal R}=0.5$~\cite{Blazey:2000qt}.
An external JES correction is determined by calibrating
the energy deposited in the jet cone  
using transverse momentum balance in exclusive photon+jet and dijet events in data~\cite{Abazov:2013hda}.
When a muon track overlaps the jet cone, twice the $p_T$ of the muon is added to the jet \pt, assuming
that the muon originates from a semileptonic decay of a hadron belonging to the jet and that the neutrino has the same $p_T$ as the muon.
In addition, we use the difference in single-particle responses between data and Monte Carlo (MC) simulation to provide a parton-flavor dependent JES correction~\cite{Abazov:2013hda}.
This correction significantly  reduces the bias in the jet energy and the total JES uncertainty of the
jets initiated by $b$ quarks. 
Jet energies  in simulated events are also corrected 
for residual differences in energy resolution and energy scale between data and simulation.
These correction factors are measured by comparing data and simulation 
in Drell$-$Yan ($\Z\to ee$) events with accompanying jets~\cite{Abazov:2013hda}.

The typical JES uncertainty is approximately $2\%$. We improve this by
calibrating the jet energy after event selection through a constant scale factor \kjes\ measured in the lepton+jets
final state using jets associated with $W$~boson decay~\cite{Abazov:2014dpa,Abazov:2015spa}. 
This  approach was first applied in Ref.~\cite{Abazov2012o}.
We apply the \kjes\ factor to the jet $p_T$ in data as 
$p^{\rm corr}_T = p_T / k_{\rm JES}$, independently for each data taking period.
We use the correction factors averaged over $e$+jets and $\mu$+jets final states (Table~\ref{tab:kjes}). 
The uncertainties related to the determination and propagation of the \kjes\ scale factor are accounted for as systematic uncertainties and described in Section \ref{sec:systematics}.
\renewcommand{\arraystretch}{1.2}
\begin{table}[!htb]
\begin{tabular}[t]{l|c c}
\hline\hline
Data taking period  & \parbox{3cm}{Integrated luminosity,~pb$^{-1}$} &  \kjes\ \\ \hline
RunIIa   &  \parbox{3cm}{1081} & 0.993 $\pm$ 0.016\\ 
RunIIb1  & \parbox{3cm}{1223} & 1.027 $\pm$ 0.013 \\ 
RunIIb2  & \parbox{3cm}{3034} & 1.033 $\pm$ 0.008 \\ 
RunIIb3   & \parbox{3cm}{ 4398} & 1.026$^{}$ $\pm$ 0.006\\ \hline\hline
\end{tabular}
\caption{
The integrated luminosity and the jet energy scale correction factor \kjes,
averaged over $e$+jets and $\mu$+jets channels~\cite{Abazov:2014dpa,Abazov:2015spa}, for the four separate data taking periods.
\label{tab:kjes}}
\end{table}

We use a multivariate analysis (MVA) technique to identify jets originating from $b$ quarks~\cite{Abazov2010k,Abazov:2013gaa}. 
The algorithm combines the information from the  impact parameters of tracks and from variables 
that characterize the properties of
secondary vertices within jets.
Jet candidates for $b$ tagging
are required to have at least two tracks with $p_T>0.5$ GeV originating from the vertex of the $p\bar{p}$ interaction,
and to be matched to a jet reconstructed from just the charged tracks. 

The missing transverse momentum,  
\met,  is reconstructed from the energy deposited in the calorimeter 
cells, and all corrections to $p_T$ for leptons and jets are propagated into a revised \met.
A significance in \met, symbolized by \metsig, 
is defined through a likelihood ratio based on the \met\ probability distribution, calculated from the expected resolution in \met\ and the energies of electrons, muons, and jets.

\subsection{Event selection}
We follow the approach developed in Ref.~\cite{Abazov:2011cq} to select dilepton events,
using the criteria listed below:
\begin{enumerate}[(i)]
\item
\label{sel:first}
For the $ee$ and $\mu\mu$ channels,
we select events that pass at least one single-lepton trigger, while for the $e\mu$ channel we 
consider events selected through 
a mixture of single and multilepton triggers and lepton+jet
triggers.
Efficiencies for single electron and muon  triggers are measured using
$\Z\to ee$ or $\Z\to\mu\mu$ data, and found to be $\approx 99$\%  and $\approx 80$\%, 
respectively, in dilepton events. 
For the $e\mu$ channel, the trigger efficiency is $\approx$ 100\%.

\item We require at least one $p\bar{p}$ interaction vertex in the interaction region with $ |z| < 60$ cm, where
$z$ is the coordinate along the beam axis,  and $z=0$ is the center of the detector.
At least three tracks with $p_T>0.5$ GeV must be associated with this vertex.

\item We require at least two isolated leptons with \mbox{$\pt>15$ GeV},
  both originating from the same interaction  vertex.
The two highest-\pt\ leptons  must have opposite electric charges. 

\item  
\label{sel:incl}
To reduce the background from bremsstrahlung in the $e\mu$ final state, we require the distance 
in ($\eta,\phi$) space between the electron and the muon trajectories 
to be ${\cal R}(e,\mu)>0.3$.

\item 
\label{sel:jets}
We require the presence of at least two jets with $\pt>20$~GeV and $|\etadet|<2.5$.

\item 
\label{sel:btag}
The $\ttbar$ final state contains two $b$ quark jets. To improve the separation between signal and background, we apply a selection using the 
$b$ quark jet identification MVA discriminant to demand that at least one of the two jets with highest $p_T$ is $b$ tagged~\cite{Abazov2010k,Abazov:2013gaa}. The $b$ tagging helps significantly in rejecting Z boson related backgrounds. We apply requirements on the MVA variable that provide $b$ quark jet identification efficiencies of 84\%  in $e\mu$, 80\% in $ee$, and 78\%  in $\mu\mu$ final states, with background misidentifications rates
of 23\%, 12\%, and 7\%, respectively.

\item
\label{sel:topo}
Additional selection criteria based on  global event  properties further improve the signal purity.
In $e\mu$ events, we require $H_T>110$~GeV,
where $H_T$ is the scalar sum of the $p_T$ of the leading lepton and the two leading jets.
In the $ee$ final state, we require \metsig$>5$,
while in the $\mu\mu$ channel, we require  \met$>40$~GeV and \metsig$>2.5$.

\item
  \label{sel:int}
  In rare cases, the numerical integration of the matrix elements described in Section~\ref{sec:ME}
   may yield extremely small probabilities that prevent us from using the event in the analysis. 
  We reject such events using a selection that has an efficiency of 99.97\%
  for  simulated \ttbar\ signal samples. For background MC events, the efficiency is 
  99.3\%. No event is removed from the final data sample because of this requirement.
\end{enumerate}

\subsection{Simulation of signal and background events}

The main sources of background in the \dilepton\  channel  are 
Drell$-$Yan production ($q\bar{q}\to (\Z\to\ell\ell)$+jets), 
 diboson production ({\sl WW, WZ,} and {\sl ZZ}), and  instrumental background.
The instrumental background arises mainly from \mbox{$(W \to \ell \nu)$+jets}  and multijet  events, in which one or two 
jets are misidentified as electrons, or where 
muons or electrons originating from  semileptonic decays of heavy-flavor hadrons appear to be isolated.
To estimate the \ttbar\ signal efficiency and the background contamination,
we use MC simulation for all contributions, except for the instrumental background, 
which is estimated from data.

The number of expected  \ttbar\ signal events is estimated using the LO matrix element generator \alpgen\ (version~v2.11)~\cite{Mangano2003}
for the hard-scattering process, with up to two additional partons,
interfaced with the \pythia\ generator~\cite{Sjostrand2006}  (version~6.409, with a D0 modified Tune~A~\cite{Affolder2002})
for parton showering and hadronization.
The CTEQ6M parton distribution functions (PDF)~\cite{Pumplin2002,Nadolsky:2008zw} 
are used in the event generation, with the top quark mass set to 172.5~GeV.
The next-to-next-to LO (NNLO) \ttbar\ cross section of $7.23^{+0.11}_{-0.20}$~pb~\cite{Czakon:2011xx} is used for the normalization.
For the calibration of the ME method, we also use events generated at  $m_{t}=$~165~GeV, 170~GeV, 175~GeV, and 180~GeV.
Those samples are simulated in the same way as the sample with the $m_{t}=172.5$~GeV. Drell$-$Yan samples are also simulated using \alpgen\ (version~2.11) 
for the hard-scattering process, with up to three additional partons, and the
\pythia\ (version~6.409, D0 modified Tune A) generator for parton showering and hadronization.
We  separately generate processes  corresponding to $Z$-boson production with heavy
flavor partons, $(Z\to \ell\ell) + b\bar{b}$ and $(Z\to \ell\ell)+  c\bar{c}$,
and light flavor partons.
Samples with light partons only are generated separately for the parton multiplicities of 0, 1, 2 and 3, 
samples with the heavy flavor partons are generated including additional 0, 1 and 2 light partons.
The MC cross sections for all Drell$-$Yan samples are scaled up with a next-to-LO (NLO) $K$-factor of 1.3,
and cross sections for heavy-flavor samples are scaled up with additional $K$-factors of 
1.52 for $(Z\to \ell\ell) +  b\bar{b}$ and  1.67 for $(Z\to \ell\ell) +  c\bar{c}$,  as estimated with the MCFM program \cite{Ellis:2006ar}.
In the simulation of diboson events, the \pythia\ generator is used for both hard scattering 
and parton showering. To simulate effects from additional overlapping \ppbar\ interactions, ``zero bias'' events 
are selected randomly in collider data and overlaid on the simulated events. Generated MC events are processed using a \geant-based~\cite{geant3} simulation of the D0 detector.

\subsection{Estimation of instrumental  background contributions}
In the $ee$ and $e\mu$ channels, we determine the contributions from events in data
with jets misidentified as electrons through the ``matrix method"~\cite{Abazov:2007kg}.
A sample of events ($n_{\text{loose}}$) is defined using
the same selections as given for \ttbar\ candidates in items (\ref{sel:first})~--~(\ref{sel:topo}) above,
but omitting the requirement on the electron MVA discriminant.
For the dielectron channel, we drop the MVA requirement on one of the randomly-chosen electrons.

Using $\Z \to ee$ data, we measure the efficiency $\varepsilon_e$
that events with electrons must pass the requirements on the electron MVA discriminant.
We measure the efficiency $f_e$  that events with no electron pass the electron MVA requirement by
using $e\mu$ events selected with criteria  (\ref{sel:first})~--~(\ref{sel:jets}),
but requiring leptons of same electric charge.
We also apply a reversed isolation requirement to the muon,
$E^{\mu,\text{iso}}_T/p^{\mu}_T > 0.2$, $p^{\mu,\text{iso}}_T/p^{\mu}_T > 0.2$,
and $\met<15$~GeV, to minimize the contribution from $W$+jets events.

We extract the number of events with misidentified electrons ($n_{f}$), and 
the number of events with true electrons ($n_e$), by solving the equations
\begin{equation}
\begin{aligned}
n_\text{{loose}}&= n_{e}/\varepsilon_e + n_{f}/f_e ,\\
n_{\text{tight}} &= n_{e} + n_{f} ,
\end{aligned}
\end{equation}
where $n_{\text{tight}}$ is the number of events remaining after implementing selections (\ref{sel:first})~--~(\ref{sel:topo}). 
The factors $f_e$ and $\varepsilon_e$ are measured for each jet multiplicity (0, 1, and 2 jets),
and separately for electron candidates in the central and end sections of the calorimeter.
Typical values of $\varepsilon_e$ are 0.7~--~0.8 in the CC and
0.65~--~0.75 in the EC. Values of $f_e$ are 0.005~--~0.010 in the CC, and
0.005~--~0.020 in the EC.

In the $e\mu$ and $\mu\mu$ channels, we determine the number of events 
with an isolated muon arising from decays of hadrons in jets by
relying on the same selection as for the $e\mu$ or $\mu\mu$ channels, but
requiring that both leptons have the same charge. 
In the $\mu\mu$ channel, the number of background events is taken to be 
the number of same-sign events. 
In  the $e\mu$ channel, 
it is the number of events in the same-sign sample after subtracting the contribution from 
events with misidentified electrons in the same way as it is done in Ref.~\cite{Abazov:2013wxa}.

To use the ME technique, we need  a pool of events to calculate probabilities corresponding to the instrumental background.
In the $e\mu$ channel, we use the loose sample defined above to model misidentified electron background.
Using this selection we obtain a background sample of 2901 events.
In the  $\mu\mu$ channel, the estimated number of multijet and $W$+jets background events is zero (Table~\ref{tab:yield}). 
In the $ee$ channel,   the number of such events is
too small to provide a representative instrumental background sample.  Instead we increase the number of background events due to $Z$-boson production by the corresponding amount in the calibration procedure.

 \subsection{Sample composition}
\begin{figure}[!htb]
\begin{minipage}[t]{0.48\textwidth}
\begin{center}
\includegraphics[width=\textwidth]{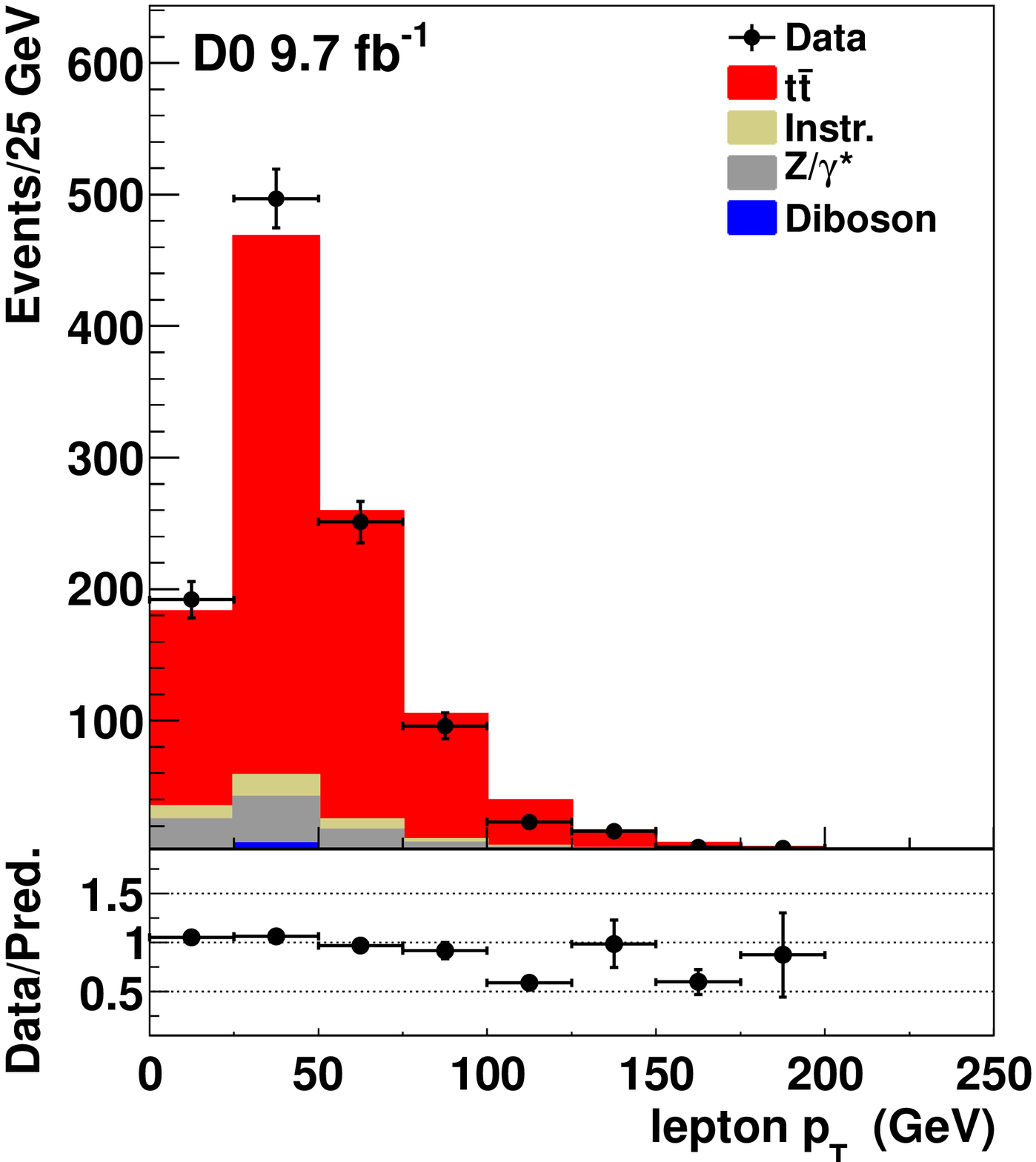}
\caption{The distributions in lepton $p_T$  and  the ratio of data to predictions for the combined $ee$,  $e\mu$, and $\mu\mu$  final states  after applying requirements (\ref{sel:first})~--~(\ref{sel:topo}).}
\label{fig:leppt_CP}
\end{center}
\end{minipage}
\hfill
\begin{minipage}[t]{0.48\textwidth}
\begin{center}
\includegraphics[width=.99\textwidth]{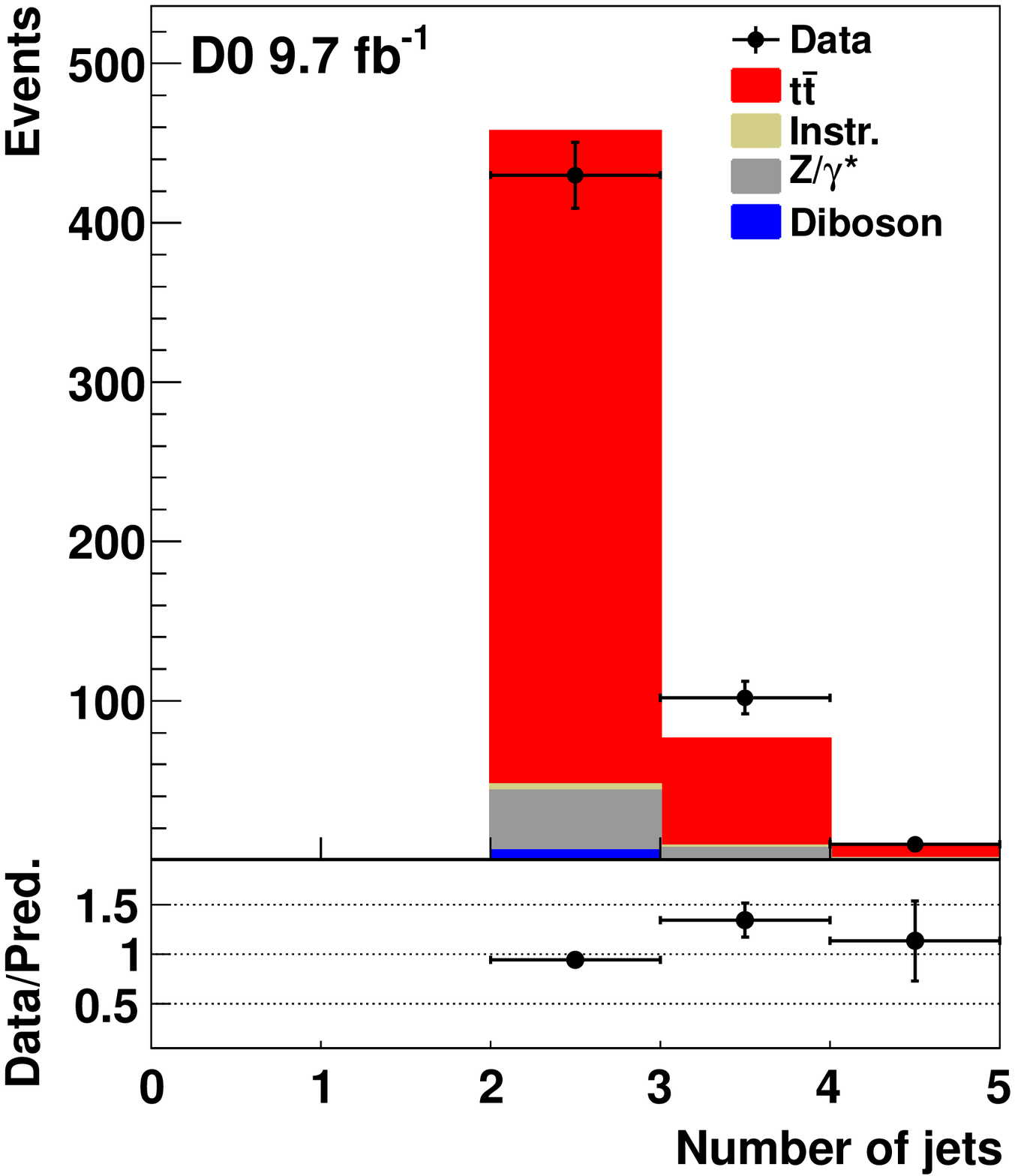}
\caption{The distributions in the number of jets  and  the ratio of data to the prediction for the combined $ee$,  $e\mu$, and $\mu\mu$  final states after applying requirements (\ref{sel:first})~--~(\ref{sel:topo}).}
\label{fig:njt_CP}
\end{center}
\end{minipage}
\end{figure}

\begin{figure}[!htb]
\begin{minipage}[t]{0.48\textwidth}
\begin{center}
\includegraphics[width=\textwidth]{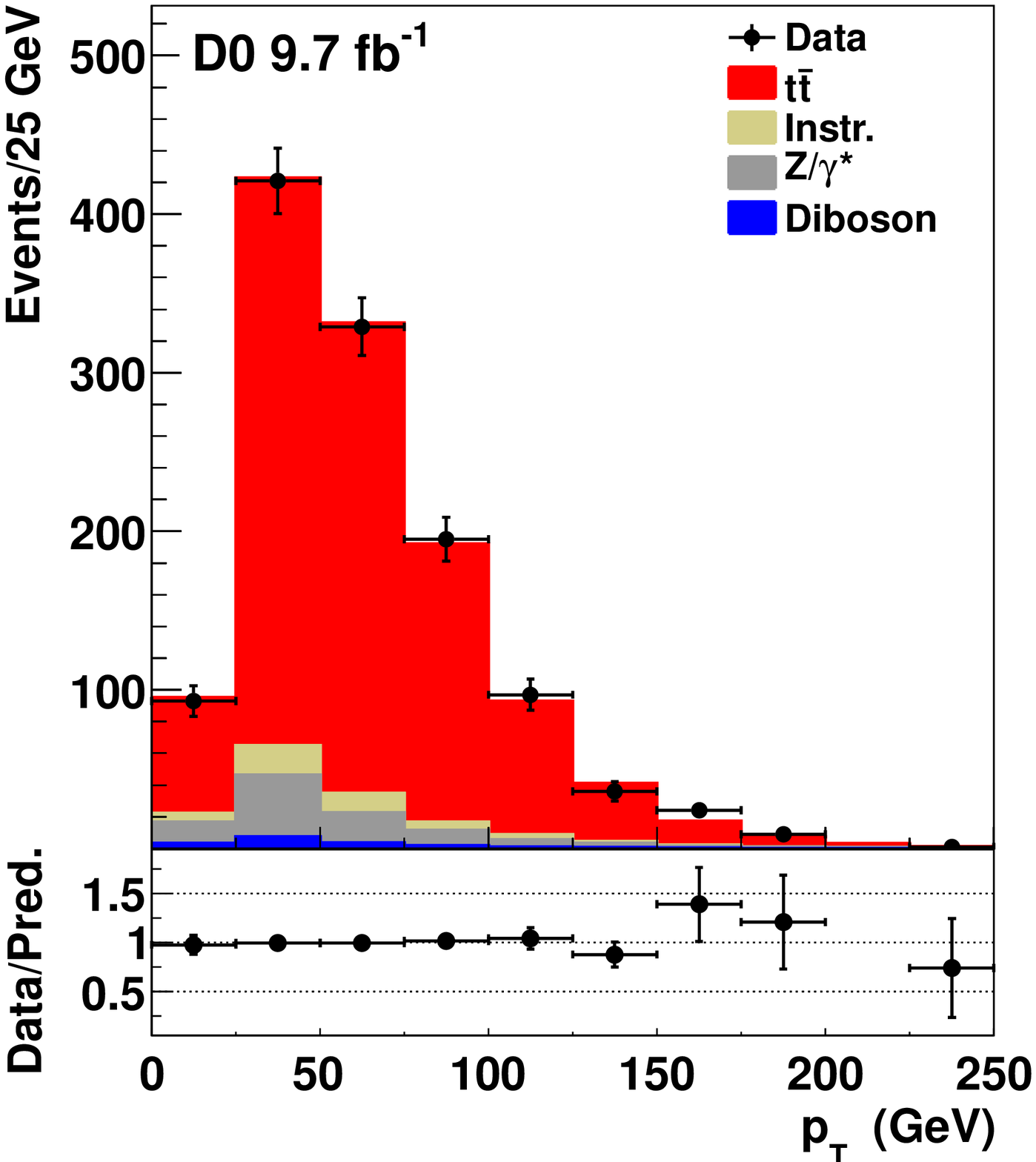}
\caption{ The distributions in  jet $p_T$ after implementing the \kjes\ correction,  and the ratio of data to the prediction for the combined $ee$,  $e\mu$, and $\mu\mu$  final states  after applying requirements (\ref{sel:first})~--~(\ref{sel:topo}).}
\label{fig:jtptkjes_CP}
\end{center}
\end{minipage}
\hfill
\begin{minipage}[t]{0.48\textwidth}
\begin{center}
\includegraphics[width=\textwidth]{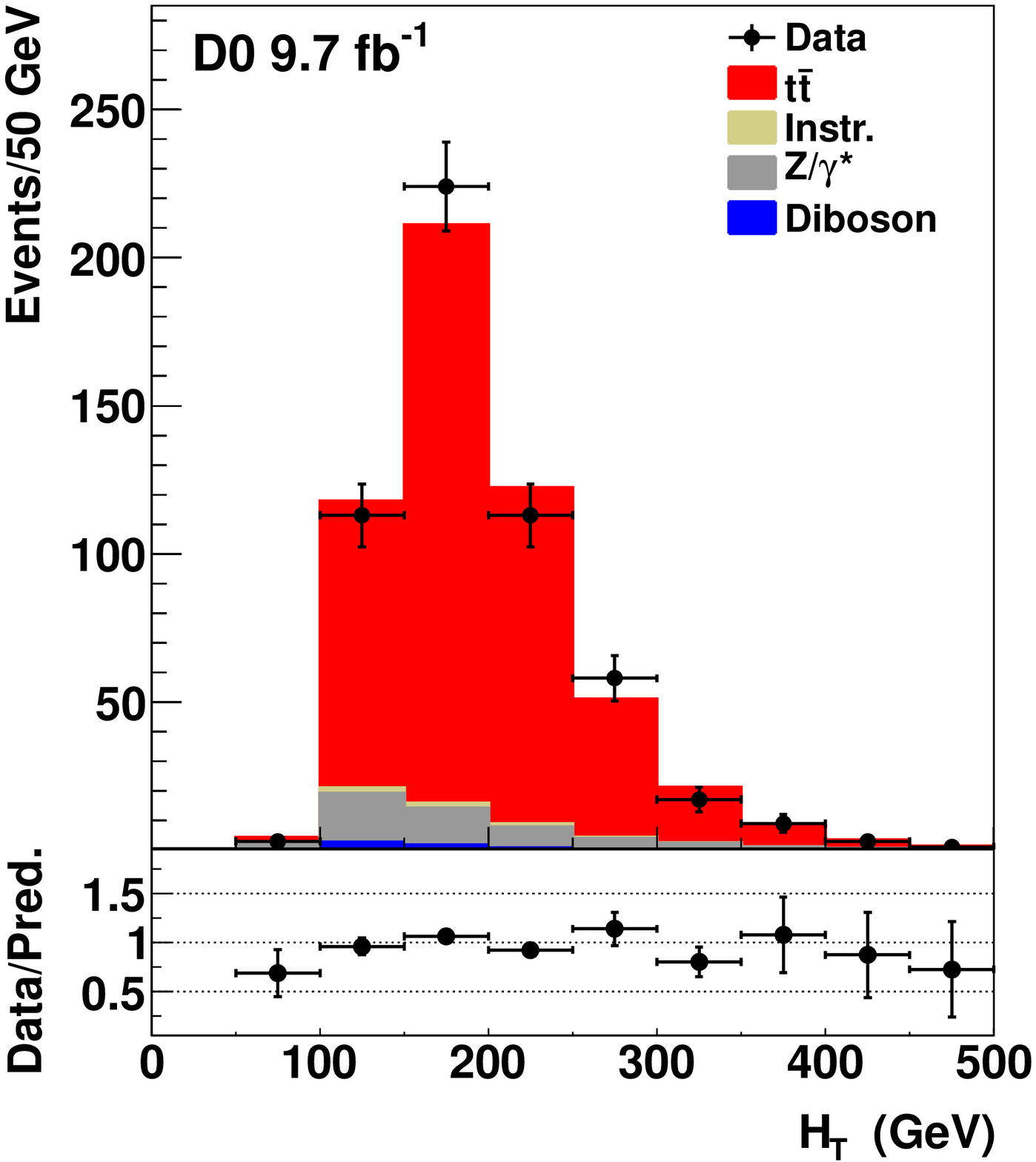}
\caption{ The distributions in $H_T$ after implementing the \kjes\ correction, and the ratio of data to the prediction for the combined $ee$,  $e\mu$, and $\mu\mu$  final states  after applying requirements (\ref{sel:first})~--~(\ref{sel:topo}).}
\label{fig:htkjes_CP}
\end{center}
\end{minipage}
\end{figure}

 The numbers of predicted background events as well as the expected numbers of signal 
events for the final selection in $e\mu$, $\mu\mu$,  and $ee$  channels are given in Table~\ref{tab:yield}. 
They show the high signal purity of the selected sample. 
The  $e\mu$ channel has a relatively low fraction of the \Z+jets background events because the electron and muon are produced 
through the cascade decay of the $\tau$-lepton, $\Z\to\tau\tau \to e\mu\nu_e\nu_\mu$.
Comparisons between distributions measured in data and predictions after the final selection are shown  in Figs.~\ref{fig:leppt_CP}-\ref{fig:htkjes_CP} for the combined $ee$, $e\mu$, and $\mu\mu$  channels. Only statistical uncertainties are shown. The predicted number of \ttbar\ and background events is normalized to the number of events found in data.
The jet $p_T$ and $H_T$ distributions in Figs.~\ref{fig:jtptkjes_CP} and \ref{fig:htkjes_CP} are  shown
after applying the \kjes\
correction from the \ljets\ analysis~\cite{Abazov:2014dpa,Abazov:2015spa}.
\begin{table}[!htbp]
\begin{tabular}[t]{c|cccc c|c}
\hline\hline

 & $Z/\gamma^\star$ + jets & Diboson & Instr. &\ttbar&\ Total\ \ & Data \\ 
$e\mu$     & $13.0^{+1.7}_{-1.6}$  & $3.7^{+0.8}_{-0.8}$  & $16.4^{+4.0}_{-4.0}$  & $260.6^{+22.5}_{-16.3}$  & $293.8^{+23.5}_{-17.7}$  & 346 \\
$ee$          & $13.8^{+2.1}_{-1.9}$  & $1.9^{+0.4}_{-0.4}$  & $1.8^{+0.2}_{-0.2}$  & $88.0^{+9.1}_{-8.2}$  & $105.5^{+10.3}_{-9.5}$  & 104 \\
$\mu\mu$ & $10.6^{+1.3}_{-1.4}$  & $1.7^{+0.4}_{-0.4}$  &   $0^{+0.05}_{-0.05}$  & $76.0^{+6.2}_{-4.1}$  & $88.3^{+6.7}_{-4.7}$  & 92  \\ \hline
$\ell\ell$     & $ 37.4 ^{+5.1}_{-4.9}$     &    $7.3  ^{+1.6}_{-1.6}$ & $18.2^{+4.0}_{-4.0}$    & $424.6^{+37.8}_{-28.6}$ & $487.6^{+40.5}_{-31.9}$  & 545 \\ 

\hline\hline
\end{tabular}
\caption{The numbers of expected background and \ttbar\ events,  and the number of events observed in data. The  NNLO cross section is used to normalize the \ttbar\ content. Systematic uncertainties are shown for all the expected numbers.
  \label{tab:yield}}
\end{table}

\section{MASS DETERMINATION Method}

\subsection{Matrix Element Technique}
\label{sec:ME}

This measurement uses the matrix element technique~\cite{Abazov:2004cs}. This method provides the most precise \mtop measurement  at the Tevatron in the
\ljets\ final state~\cite{Abazov:2014dpa,Abazov:2015spa}, and  was applied
in previous measurement of \mtop in the dilepton final state using 5.3~fb$^{-1}$ of integrated luminosity~\cite{Abazov2011c}.
The ME method used in this analysis is described below.

\subsection{Event probability calculation}
\label{sec:probability}
The ME technique  assigns  a probability to each event,  which is calculated as
\begin{equation}
P(x,f_{t\bar{t}},m_t) = f_{t\bar{t}}\cdot P_{t\bar{t}}(x,m_t) + (1-f_{t\bar{t}}) \cdot P_{\rm bkg}(x),
\label{eq:event_prob}
\end{equation} 
where $f_{t\bar{t}}$ is the fraction of \ttbar\ events in the data, and
$P_{t\bar{t}}$ and $P_{\rm bkg}$ are 
the respective per-event probabilities calculated under the hypothesis that the selected event is either a \ttbar\
event, characterized by a  top quark mass $m_t$, or background.
Here, $x$ represents the set of measured observables, i.e., $p_T$, $\eta$, and $\phi$ for jets and leptons.
We assume that the masses of top quarks and anti-top quarks are the same.
The probability $P_{t\bar{t}}(x,m_t)$ is calculated as 
\begin{equation}
\label{eq:prob_integ}
\begin{split}
P_{t\bar{t}}(x,m_t) = \frac{1}{\sigma_{\rm obs}(m_t)} \int f_{\rm PDF}(q_1) f_{\rm PDF}(q_2)\times \\
\times\frac{(2\pi)^4 |\mathscr{M}(y,m_t)|^2}{q_1 q_2 s} W(x,y) d\Phi_6 dq_1 dq_2,
\end{split}
\end{equation}
where $q_1$ and $q_2$ represent the respective fractions of proton and antiproton momenta carried by the initial state partons,
$f_{{\rm PDF}}$ represents the  parton distribution functions, 
and $y$ refers to partonic four-momenta of the final-state objects.
The detector transfer functions (TF), $W(x,y)$, correspond to the probability for reconstructing parton 
four-momenta $y$ as the final-state observables  $x$. The term $d\Phi_6$ represents the six-body  phase space, and 
$\sigma_{\rm obs}(m_t)$ is the  \ttbar\  cross section observed at the reconstruction level, 
calculated using the matrix element $\mathscr{M}(y,m_t)$, corrected for selection efficiency.
The LO matrix element $\mathscr{M}(y,m_t)$ for the processes 
$q\bar{q} \to t\bar{t} \to W^+W^- b\bar{b} \to \ell^+\ell^-\nu_\ell \bar{\nu_\ell} b\bar{b}$ is used in our calculation~\cite{Mahlon:1997uc} and it contains a Breit-Wigner function to represent each $W$ boson and top quark mass.
The matrix element is averaged over the colors and spins of the initial state partons,
and summed over the  colors and spins of the final state partons.
The $gg$ matrix element is neglected, since it comprises only 15\% of the total \ttbar\ production cross-section at the Tevatron. Including it does not significantly improve the statistical sensitivity of the method.

The electron momenta and the directions of all
reconstructed objects are assumed to be perfectly measured and are
therefore represented through $\delta$ functions, $\delta(x-y)$, reducing thereby the dimensionality
 of the integration.  This leaves the magnitues of the jet and muon momenta to be modelled.
Following the same approach as in the previous measurement~\cite{Abazov2011c},
we parametrize the jet energy resolution by a sum of 
two Gaussian functions with parameters depending linearly 
on parton energies, while the resolution in the curvature of the muon
($1/p^{\mu}_T$) is described by a single Gaussian function.
All TF parameters are determined from simulated
\ttbar\ events.
We use the same parametrizations for the transfer functions as in the \ljets\ \mtop\ measurement.
The detailed description of the TFs is given in Ref.~\cite{Abazov:2015spa}.

The masses of the six final state particles are set to 0 except for the $b$
quark jets, for which a mass of 4.7 GeV is used.
We  integrate over 8 dimensions in the $ee$ channel, 9 in the $e\mu$ channel,
and 10 in the $\mu\mu$ channel.
As integration variables  we use  the 
top and antitop quark masses, the $W^+$ and $W^-$ boson masses, the transverse momenta of the two jets, the \pt\ and $\phi$ of the \ttbar\ system, and $1/p^{\mu}_T$ for muons.
This choice of variables differs from that of
the previous measurement~\cite{Abazov2011c}, providing a  factor of $\approx 100$ reduction in integration time.

To reconstruct the masses of the top quarks and  $W$~bosons,  we solve the kinematic equations
analytically by summing over the two possible jet-parton assignments
and over all real solutions for each neutrino momentum~\cite{Fiedler:2010sg}.
If more than two jets exist in the event, we use only the two  with highest transverse momenta.
The integration is performed using the MC based 
numerical integration algorithm  VEGAS~\cite{Lepage1978,Lepage1980}, 
as implemented in the GNU Scientific Library~\cite{Galassi2009}.

Since the dominant source of background in the dilepton final state is from \zjets events, as can be seen from Table~\ref{tab:yield},  we consider only the \zjets\ matrix element in the calculation of  the background probability, $P_{\rm bkg}(x)$.
The LO $(\Z\to \ell\ell)$+2\,jets ME from the \vecbos\ generator~\cite{Berends:1990ax} is used in this analysis.
In the $e\mu$ channel, background events are produced through the  $(\Z\to \tau\tau \to \ell\ell)$+2\,jets 
processes. Since $\Z\to \tau\tau$ decays are not implemented in \vecbos, we use
an additional transfer function to describe the energy of the final state lepton relative 
to the initial $\tau$ lepton, obtained from parton-level
information~\cite{Fiedler:2010sg}.
As for $P_{t\bar{t}}(x,m_t)$,
the directions  of the jets and charged leptons are assumed to be well-measured, and each kinematic
solution is weighted according to the \pt\ of the \zjets\ system.
The integration of the probability $P_{\rm bkg}(x)$  is
performed over the energies of the two partons initiating the selected jets and both possible assignments of jets to top quark decays.

The normalization of the background  per-event probability could be defined in the same way as for the signal probabilities,  i.e. by dividing the probabilities by $\sigma_{\rm obs}$.
However, the calculation of the integral equivalent to  Eq.~(\ref{eq:prob_integ}) for the background
requires significant computational resources, and therefore a different approach is chosen.
We use a large ensemble including \ttbar\ and background events in known proportion. We fit the fraction of background events in the ensemble by adjusting the background normalization. The value which minimizes the  the difference between the fitted signal fraction and the true one is chosen as the background normalization factor (see~ Ref.~\cite{Grohsjean:2008zz} for more details).
\subsection{Likelihood evaluation and $m_t$ extraction}
\label{sec:likelihood}
To extract the top quark mass from a set of $n$ events with
measured observables $x_1,..,x_n$, we construct a log-likelihood function from the
event probabilities
\begin{equation}
\label{eq:neg-lhood-fnc}
-\ln L(x_1,..,x_n;f_{t\bar{t}},m_t) = - \sum_{i=1}^{n}\ln (P_{\rm evt}(x_i;f_{t\bar{t}},m_t)).
\end{equation}
This function is minimized with respect to the two free parameters $f_{t\bar{t}}$
and  $m_t$.
To calculate the signal probabilities, we use  step sizes of
 2.5~GeV for $m_t$ and 0.004 for $f_{t\bar{t}}$. 
The minimum value of the log-likelihood function, $m_{\rm lhood}$, is fitted using a  second 
degree polynomial function, in which $f_{t\overline t}$ is fixed at its fitted value.
The statistical uncertainty on the top quark mass, $\sigma_{\rm lhood}$, is given by the difference
in the mass at $- \ln L_{\rm min}$ and at $- \ln L_{\rm min} + 0.5$. The  $m_t$  extractions are done separately for $ee$,  $e\mu$, and $\mu\mu$  final states and for the combination of all three channels.

\subsection{Method calibration}
\label{sec:calibration}
We calibrate the method to correct for biases in the measured mass 
and statistical uncertainty through an ensemble testing technique.
We generate  data-like ensembles with 
simulated signal and background events,
measure the top quark mass $m^i_{\rm lhood}$ 
and its uncertainty $\sigma^i_{\rm lhood}$ in  each ensemble $i$ through the minimization of the
log-likelihood function, and calculate the following quantities:
\begin{enumerate}[(i)]
\item The mean value $m_{\rm mean}$ of the $m^i_{\rm lhood}$ distribution.
Comparing  $m_{\rm mean}$ with the input in the simulation determines the bias in \mtop.

\item The mean value $\Delta m_t$ of the uncertainty distribution in $\sigma^i_{\rm lhood}$. 
This quantity characterizes the  expected uncertainty in the measured top quark mass.

\item The standard deviation of the distribution of the pull variable, $w_{\rm pull}$, or pull width, where the
pull variable is defined as $w_{\rm pull}=(m^i_{\rm lhood} - m_{\rm mean})/\sigma^i_{\rm lhood}$,
provides a correction to the  statistical uncertainty $\sigma_{\rm lhood}$.
\end{enumerate}
 We use resampling (multiple uses of a given event) when generating the ensembles.
In the D0 MC simulation, a statistical weight $w_j$  is associated with each event $j$, which is given by the product
of the MC cross section weight, 
 simulation-to-data efficiency corrections and other simulation-to-data correction factors.
The probability for an event to be used in the ensemble is proportional to its weight $w_j$.
Multiple use of the events significantly reduces the uncertainty of the ensemble testing procedure
for a fixed number of ensembles,
but leads to the overestimation of the statistical precision, for which we account through a dedicated correction factor.

\begin{figure}[!htb]
\begin{center}
\includegraphics[width=0.39\textwidth]{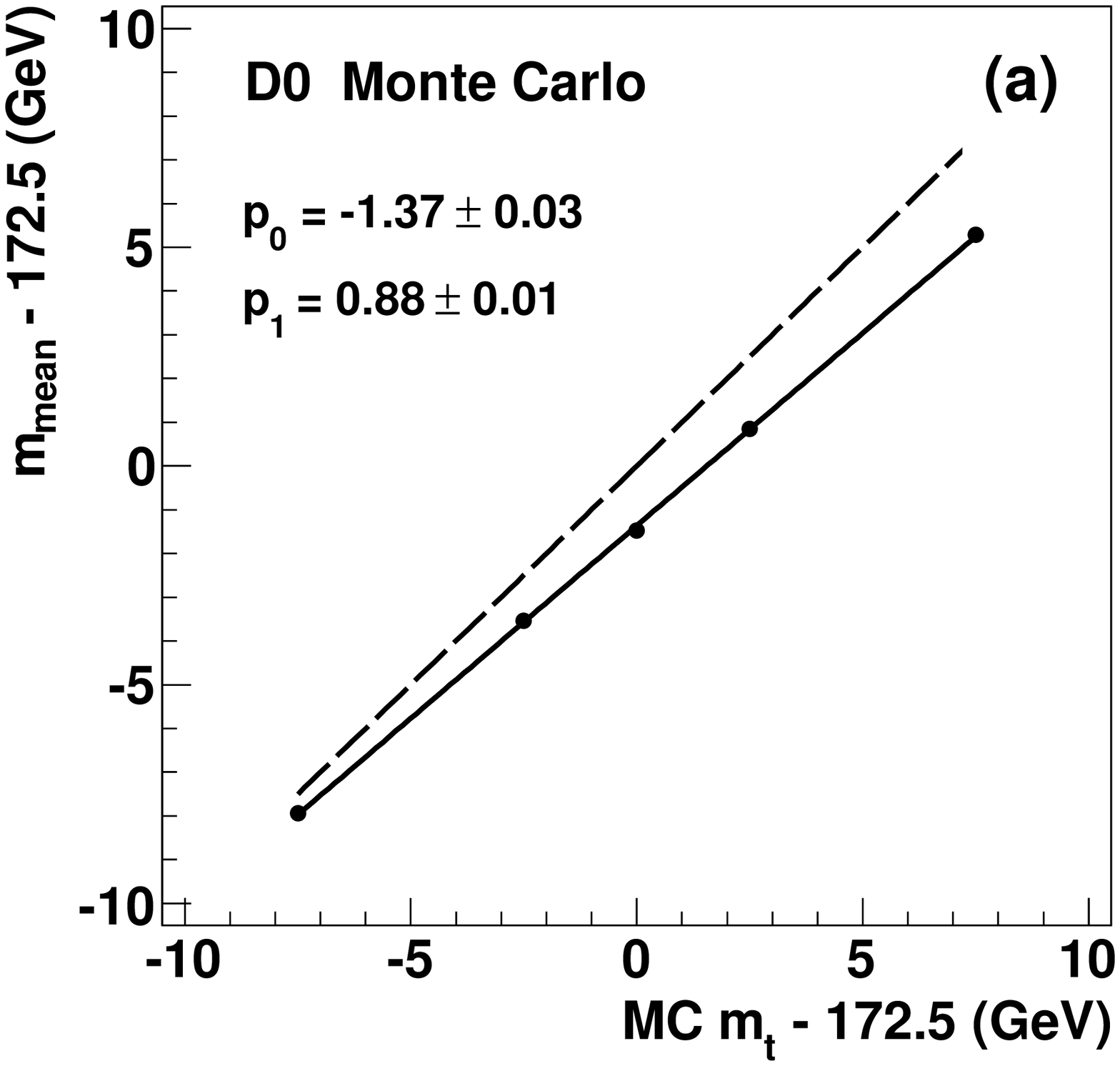}
\includegraphics[width=0.39\textwidth]{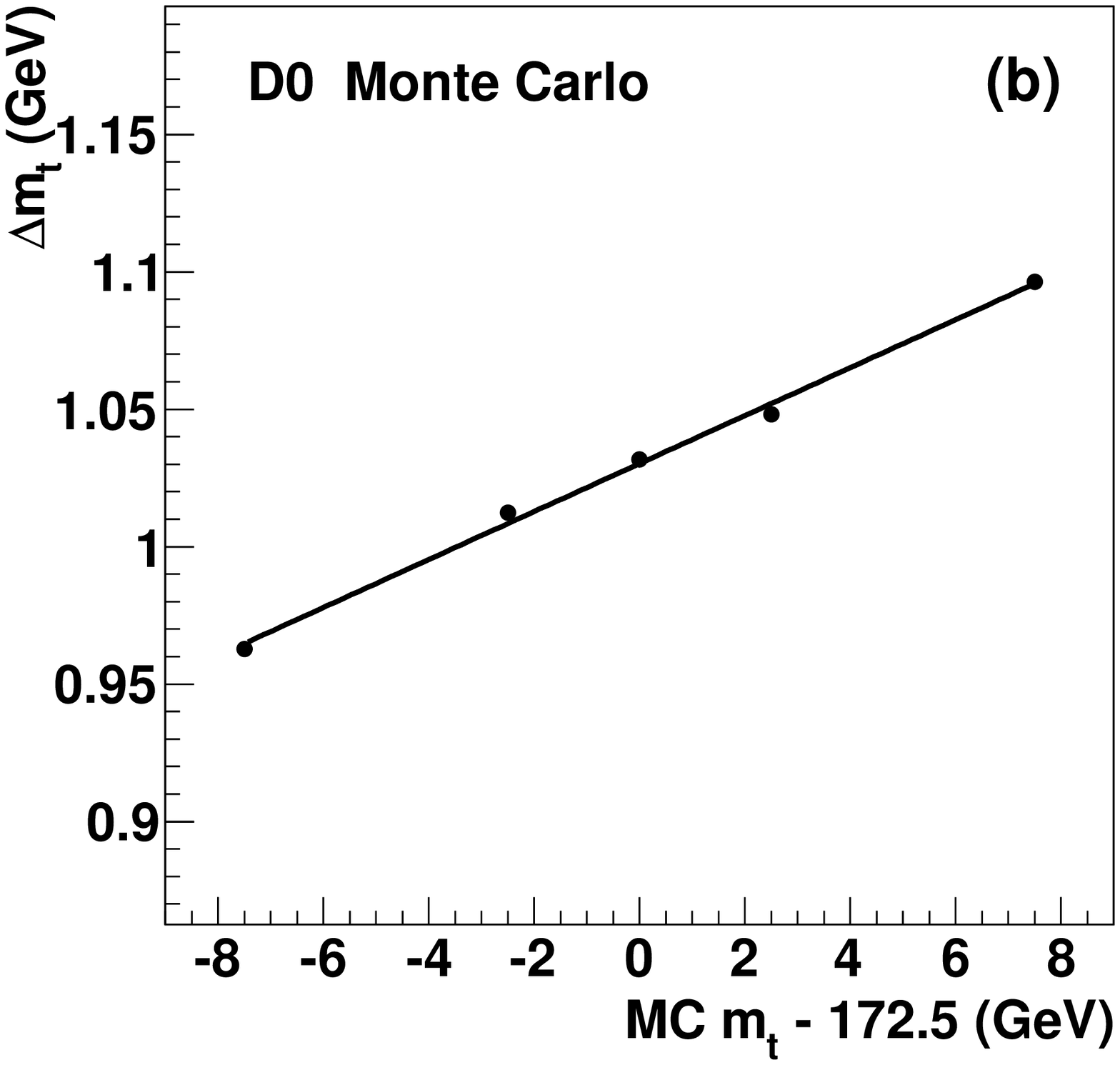}
\includegraphics[width=0.39\textwidth]{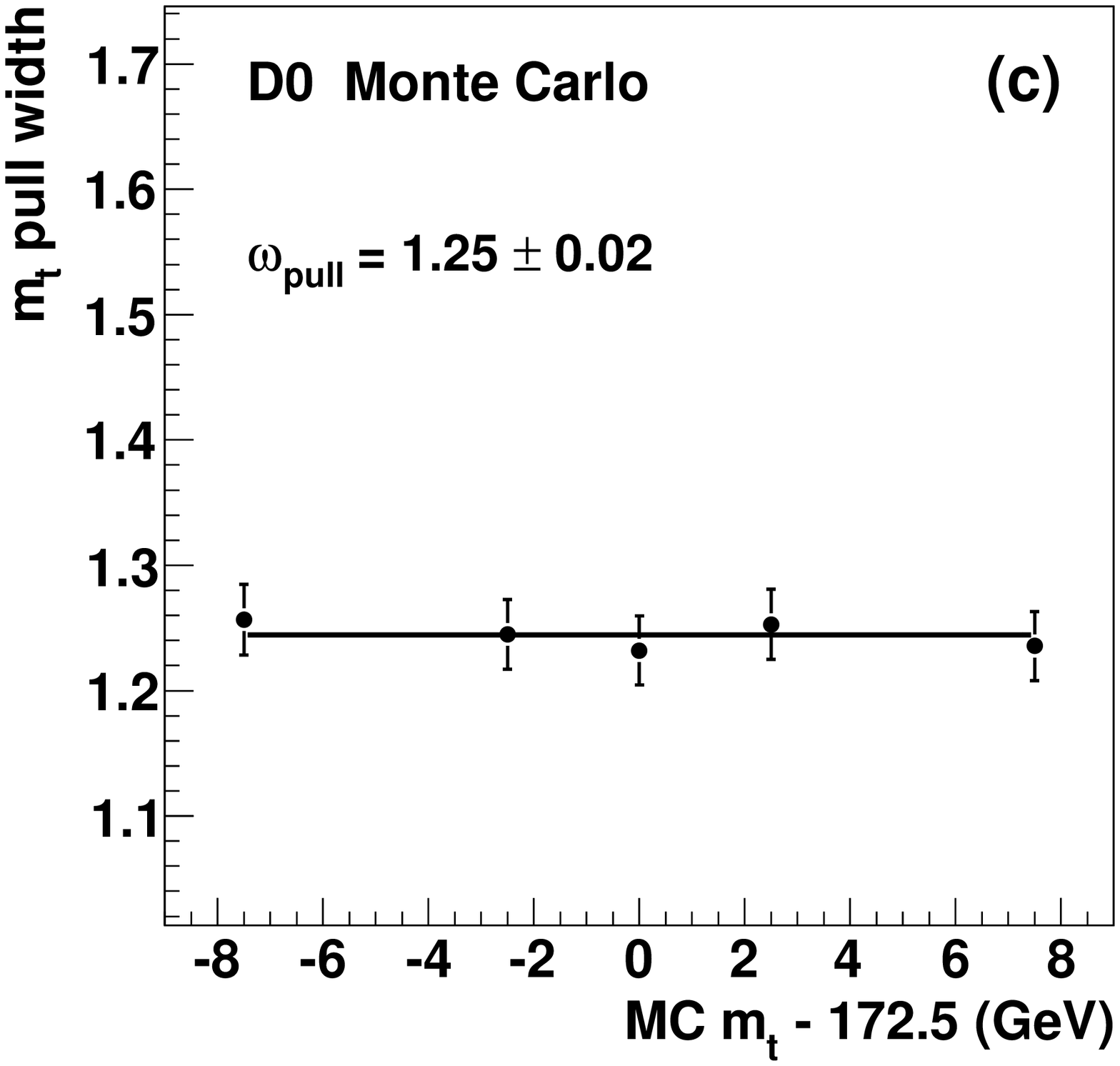}
\caption{The  response of the ME method in (a)  $m_t$, 
(b) statistical uncertainty on the  $m_t$, and (c) the pull width, shown as  a function of the MC input  $m_t$ for the combined $ee$, $e\mu$, and $\mu\mu$ channels. The error bars in (a) and (b) are invisibly small. The dashed line in (a) represents the case of ideal response.}
\label{fig:calib}
\end{center}
\end{figure}

We use 1000 ensembles per MC input mass $m_{t}$, with the number of events per ensemble equal to the number of events selected in data.
In each ensemble, the number of events from each background source is generated  following multinomial statistics, 
using the expected number of background events in Table~\ref{tab:yield}.
The number of  \ttbar\ events is calculated as the difference between the total number of events in the ensemble and
the generated number of background events. We combine all three channels to construct  a joint calibration curve.
Using MC samples  generated at five MC $m_{t}$, 
we determine a  linear calibration between  the measured and generated masses:
 $m_{\rm mean}-172.5$~GeV~$= p_0 + p_1 ({\rm MC}\ m_{t} -172.5)$~GeV. 
The relations obtained for the combination of the $e\mu$, $ee$, and $\mu\mu$ final states
are shown in Fig.~\ref{fig:calib}. The difference of the calibration curve from the ideal case demonstrates that the method suffers from some biases.

\begin{table}[!htb]
\begin{center}
\begin{tabular}[t]{l|ccc|c}
\hline\hline
Final state & $ee$   & $e\mu$ & $\mu\mu$  & \dilepton  \\
Uncertainty, GeV  & 3.69 & 1.71  & 3.57 & 1.45 \\
\hline\hline
\end{tabular}

\caption{The expected statistical uncertainties for a generated $\mtop=172.5$~GeV for the $ee$, $e\mu$, and $\mu\mu$ channels and their combination.
\label{tab:exp_uncert_chan}}
\end{center}
\end{table} 
The expected statistical uncertainty for the generated top quark mass  of 172.5 GeV is calculated as 
$\Delta m_t^{\rm exp} = \Delta m_t (172.5\ {\rm GeV}) \cdot w_{\rm pull} / p_1$, and 
given in Table~\ref{tab:exp_uncert_chan}.
\section{Fit to Data}
\label{sec:app2data}
\begin{figure}[!htb]

\begin{minipage}[t]{0.49\textwidth}
\begin{center}
\includegraphics[width=.9\textwidth]{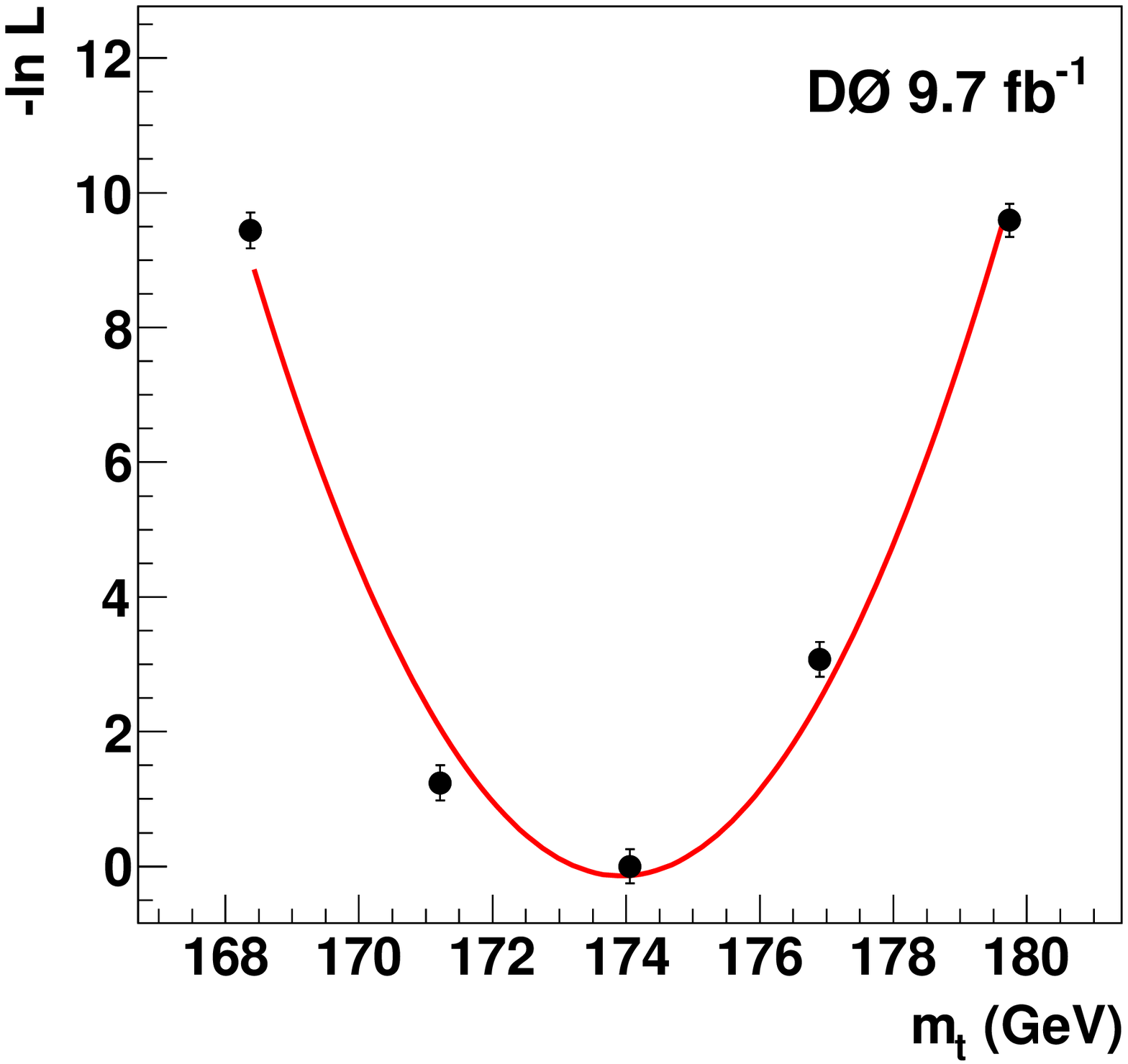}
\caption{The negative log-likelihood ratio  for the combined $ee$, $e\mu$, and $\mu\mu$ data after calibration, as a function of the input MC \mtop. The curve is the best fit to a parabola  in the interval  168.4~$-$~179.7~GeV.}
\label{fig:data_lhood}
\end{center}
\end{minipage}
\hfill
\begin{minipage}[t]{0.49\textwidth}
\begin{center}
\includegraphics[width=.9\textwidth]{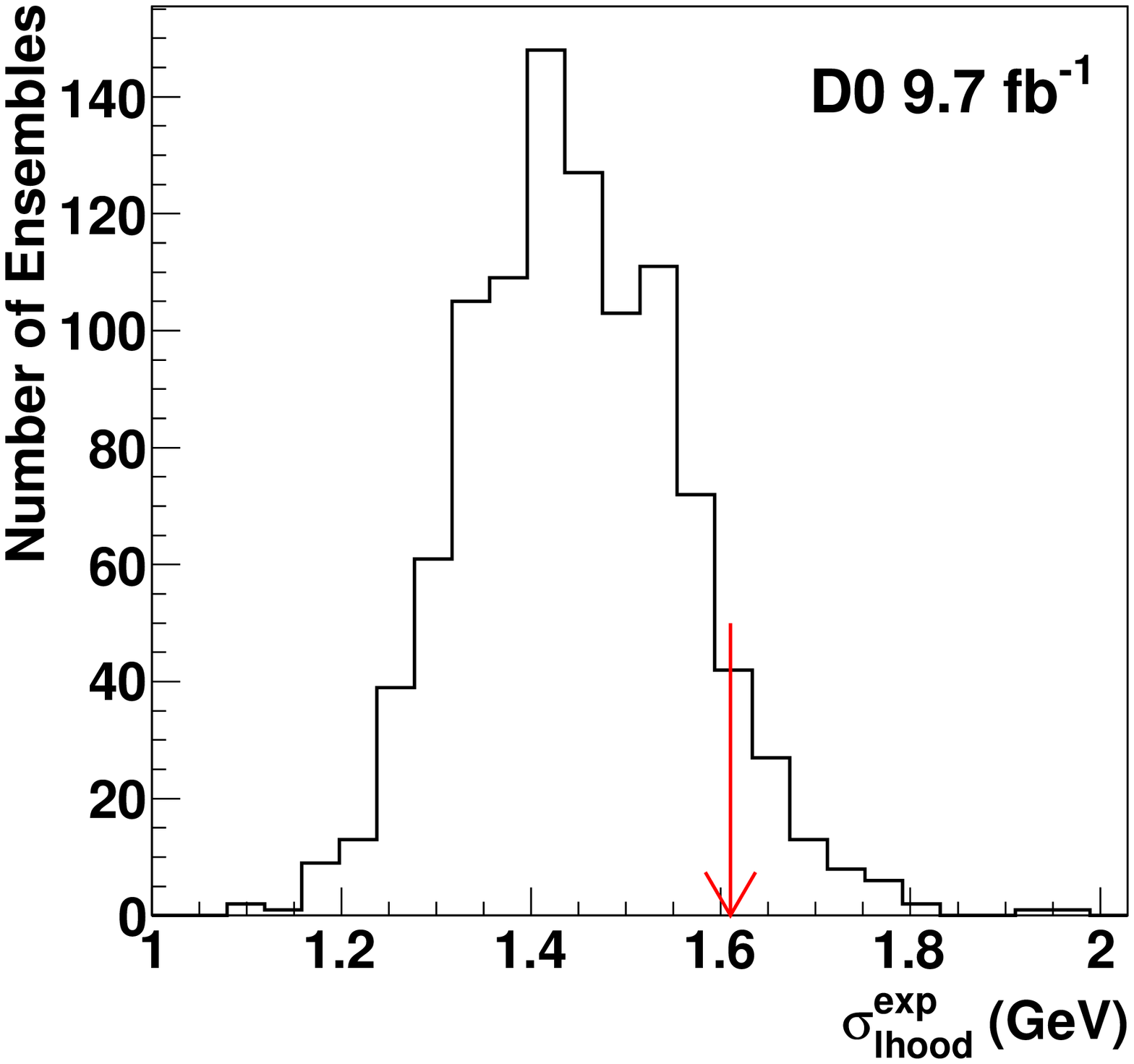}
\caption{The distribution in the expected statistical uncertainty $\sigma_{\rm lhood}^{\rm exp}$ 
for the combined  $ee$, $e\mu$, and $\mu\mu$ channels after applying calibration, for the MC input $\mtop=175$~GeV. 
The arrow indicates the statistical uncertainty for data after the calibration ($\sigma_{\rm meas}$).}
\label{fig:MCdata_uncert}
\end{center}
\end{minipage}
\end{figure}

The fit to data is first performed using an unknown random offset in the measured mass. 
 This offset is removed only after the final validation of the methodology. 
We apply the ME technique to data as follows:
\begin{enumerate}[(i)]
\item The \kjes\  correction factor from the lepton+jets mass 
analysis~\cite{Abazov:2014dpa,Abazov:2015spa} is applied to the jet $p_T$ in data as 
$p^{\rm corr}_T = p_T / k_{\rm JES}$ (Section~\ref{sec:detector}).
The uncertainties related 
to the propagation of this correction from \ljets\ to the dilepton final state are included in the systematic
uncertainties as a residual JES uncertainty and statistical uncertainty on \kjes\ scale factor discussed
in  Section~\ref{sec:JESsyst}.

\item The calibration correction from Fig.~\ref{fig:calib} is applied to $m_{\rm lhood}$ 
and $\sigma_{\rm lhood}$ 
to obtain the measured values:
\begin{equation}
\begin{aligned}
m_{\rm meas} &= (m_{\rm lhood}-p_0 - 172.5)/p_1 + 172.5~{\rm (GeV)},\\
\sigma_{\rm meas}& = \sigma_{\rm lhood}\cdot w_{\rm pull}/p_1.
\end{aligned}
\end{equation}
\item The fit to the log-likelihood function is the best fit to a parabola in an interval containing a 
10~GeV range in MC \mtop around the minimum before its calibration.
\end{enumerate} 
The log-likelihood function in data is shown in Fig.~\ref{fig:data_lhood}.
Table~\ref{tab:results_chan} shows the results for each channel separately and for their combination.
The distribution in the expected statistical uncertainty for an input MC top quark mass of 175~GeV (the closest input value to the mass obtained in data) for the three combined channels is shown in Fig.~\ref{fig:MCdata_uncert}. 

\begin{table}[!htb]
\begin{tabular}[t]{c|c}
\hline\hline
Final state & Mass (GeV) \\
\hline
$ee$  & 176.94 $\pm$ 4.65 \\
$e\mu$ & 172.18 $\pm$ 1.95 \\
$\mu\mu$ &  176.04 $\pm$ 4.82\\
\dilepton & \mmeas\ $\pm$ \mstat \\
\hline\hline
\end{tabular}
\caption{The calibrated top quark mass  for the $ee$, $e\mu$, and $\mu\mu$ channels, and for 
their combination. The quoted uncertainties are statistical.
\label{tab:results_chan}}
\end{table}

\section{Systematic Uncertainties and Results} 
\label{sec:systematics}
 Systematic uncertainties affect the measured \mtop\ in two ways. 
First, the distribution in the signal and background  
log-likelihood functions can be affected directly by a change in some parameter, leading to a bias in the calibration.
Second, the signal-to-background ratio in the selected data
can be affected by the parameter change, leading to a difference in the combined signal and background log-likelihood function, again causing a a bias in the calibration.
Ideally, these two contributions
can be treated coherently for each source of systematic uncertainty, but since the second effect
is much smaller than the first for the most important systematic uncertainties, we keep the same signal-to-background ratio in pseudo-experiments, 
except for the systematic uncertainty in the signal fraction. 
Background events are included in the evaluation of all sources of systematic uncertainty, and all systematic uncertainties are evaluated using the simulated events
with a top quark mass of 172.5~GeV.

\subsection{Systematic uncertainties in modeling signal and background}
\label{sec:SigModsyst}
We determine uncertainties related to  signal modeling 
by comparing simulations with different generators and parameters, as described below.

\textbf{Higher order corrections.}
By default, we use LO \alpgen\ to model signal events. 
To evaluate the effect of higher-order corrections on the top quark mass,  we use signal events generated  with the NLO MC
generator \mcatnlo~(version~3.4)~\cite{Frixione:2002ik,Frixione:2008ym},
interfaced to \herwig~(version~6.510)~\cite{Corcella:2000bw} for parton
showering and hadronization. The CTEQ6M PDFs~\cite{Pumplin2002,Nadolsky:2008zw}
are used to generate events at a top quark mass of $m_t=172.5$~GeV. 
Because \mcatnlo\ is interfaced to \herwig\ for simulating the showering contributions to the process of interest, we use \alpgen+\herwig\ events for this comparison, in order to avoid double-counting an uncertainty due to a different showering model.

\textbf{Initial state radiation (ISR) and final-state radiation (FSR).}
This systematic uncertainty is evaluated comparing the result
using \alpgen+\pythia\  by changing the factorization and renormalization scale parameters, up and down by a factor of 2, as done in Ref.~\cite{Abazov:2015spa}. 

\textbf{Hadronization and underlying event.}
The systematic uncertainty due to the hadronization and the underlying 
event (UE) is estimated as the 
difference between \mtop\ measured using the default \alpgen+\pythia\ events
and events generated using different hadronization models. 
We consider three  alternatives: \alpgen+\herwig, 
\alpgen+\pythia\ using Perugia Tune 2011C (with color reconnection), or using
Perugia Tune 2011NOCR (without color reconnection)~\cite{Skands2010}.
We take the largest  of these differences, which is the difference relative to  \alpgen+\herwig,  as an estimate of the systematic uncertainty for choice of effects from the hadronization and the UE.

\textbf{Color reconnection.}
We estimate the effect  of the model for color reconnection (CR) by comparing the top quark mass measured with \alpgen+\pythia\ Perugia Tune 2011C (with color reconnection), 
and with Perugia Tune 2011NOCR (without color reconnection)~\cite{Skands2010}. 
Our default \alpgen+\pythia\ tune does not have explicit CR modeling, so we consider Perugia2011NOCR as the default in this comparision.

\textbf{Uncertainty in modeling $\bm{b}$ quark fragmentation ($\bm{b}$ quark jet modeling).}
Uncertainties in simulation of $b$ quark fragmentation can affect the \mtop measurement 
through $b$ quark jet identification or transfer functions. 
This is studied using the procedure described in Ref.~\cite{Peters2006} by reweighting  $b$ quark fragmentation to match a Bowler scheme
tuned to either LEP or SLD data.

\textbf{PDF uncertainties.}
The systematic uncertainty due to the choice of PDF is estimated by
changing the 20 eigenvalues of the CTEQ6.1M PDF within their uncertainties in \ttbar\  MC simulations. 
Ensemble tests are repeated for each of these changes and the total uncertainty is evaluated 
as in Ref.~\cite{Abazov:2015spa}.

\textbf{Transverse momentum of the $\bm{t\bar{t}}$ system}. To evaluate this systematic uncertainty, 
we reconstruct the  \ttbar\ $p_{T}$ from the two leading jets, two leading leptons, and \met.
The distribution in the MC events is reweighted to match that in data using a linear fit to the $p_{T}$ distribution of the \ttbar\  system.  
To improve statistics, we combine all the dilepton channels for the extraction of the reweighting function. 

 \textbf{Heavy-flavor scale factor}. In the \alpgen\ \zlljets background samples, the fraction
of heavy-flavor events is not well modelled. 
Therefore, a heavy-flavor scale factor is applied to the $(Z\to\ell\ell)+ b\bar{b}$ and 
$(\Z\to\ell\ell)+ c\bar{c}$ cross sections
to increase the heavy-flavor content. This scale factor has an uncertainty of $\pm 20\%$. We estimate its systematic effect by changing the scale factor within this uncertainty.

\textbf{Multiple $\bm{p\bar{p}}$ interactions}. Several independent \ppbar\ interactions in the same bunch crossing may influence the measurement of \mtop. 
We reweight the number of interactions in simulated MC samples to the number of interactions found in data
before implementing any selection requirements.
To estimate the effect from a possible mismatch in luminosity profiles, we 
examine the distribution in instantaneous luminosity in both data and MC after event  selection, and reweight the instantaneous luminosity profile in MC events to match data.

\subsection{JES systematic uncertainties}
\label{sec:JESsyst}
The relative difference between the JES in data and MC simulations is described by the \kjes\ factor extracted in the \ljets\ mass measurement~\cite{Abazov:2014dpa,Abazov:2015spa}.
As mentioned above, we apply this scale factor to jet $p_T$ in data. 
In the previous dilepton analysis~\cite{Abazov2011c}, the JES and the ratio of $b$ and light jet responses were the dominant systematic uncertainties. 
The improvements made in the jet calibration~\cite{Abazov:2013hda} 
and use of the \kjes\ factor
in the dilepton channel reduce the uncertainty related to the JES from 1.5 GeV to 0.5 GeV.

\textbf{Residual uncertainty in JES}.
This uncertainty arises from the fact that the JES depends on the \pt\ and $\eta$ of the jet. The JES correction in  the \ljets\ measurement assumes
a constant scale factor, i.e., we correct the average JES, but not the \pt\ and $\eta$
dependence.
In addition, the \kjes\ correction can be affected by the different jet $p_T$  requirements 
on jets in the \ljets\ and in dilepton final states.
There can also be a different JES offset correction  due to different jet multiplicities.
We estimate these uncertainties as  follows.
We use MC events in which the jet energies are shifted upward by one standard deviation 
of the $\gamma$+jet JES uncertainty
and correct jet $p_T$ in these samples to
$p^{\rm corr MC}_T = p^{\rm MC}_T \cdot k^{\rm UP}_{\rm JES} / k_{\rm JES}$, 
where $ k^{\rm UP}_{\rm JES}$ is the JES correction  measured in the \ljets\ analysis for the MC events that are shifted up by one standard deviation. 
The $ 1/ k_{\rm JES}$ factor appears because the \kjes\  is applied to the  data and
not to MC samples.
Following the same approach as in~\cite{Abazov:2013hda},  we assume that the downward change for the JES samples has the same effect as the upward changes in jet $p_T$.

\textbf{Uncertainty on the $\bm{k_{\bf{\rm JES}}}$ factor}.
The statistical uncertainty on the \kjes\  scale factor is  0.5\% -- 1.5\% depending on the data taking period
(Table~\ref{tab:kjes}).
We recalculate the mass measured in MC with the \kjes\ correction shifted by one standard deviation.
This procedure is applied separately for each data taking period, and the uncertainties 
are summed in quadrature.

\textbf{Ratio of $\bm{b}$ and light jet responses or flavor-dependent uncertainty}.
The JES calibration used in this measurement contains a flavor-dependent  jet response correction, 
which  accounts for the difference in detector response to different jet flavors, in particular $b$ quark
jets versus light-quark jets.
This correction is applied to the jets in MC simulation through a convolution of the corrections 
for all simulated particles associated to the jet as a function of particle $p_T$ and $\eta$.
It is constructed in a way that preserves the flavor-averaged JES corrections for $\gamma$+ jets events ~\cite{Abazov:2013hda}. 
The \kjes\ correction does not improve this calibration, because it is measured in light jet flavor from
$WW \to qq^\prime$ decays.
To propagate the effect of the uncertainty  to the  measured \mtop\ value, 
we change the corresponding correction by the size of the uncertainty and recalculate \mtop.

\subsection{Object reconstruction and identification}
\label{sec:RecIDsyst}
\textbf{Trigger}.  To evaluate the impact of the trigger on our analysis, we scale the number of background events according to the uncertainty on the trigger efficiency for different channels.
The number of signal \ttbar\ events is recalculated as the difference between the number of events in data and the expected number of background events. We reconstruct ensembles according to the varied event fractions and extract the new mass.

\textbf{Electron momentum scale and resolution}.
This uncertainty reflects the difference in the absolute lepton momentum 
measurement and the simulated resolution~\cite{Abazov:2013tha} between data and MC events.
We estimate this uncertainty by changing the corresponding parameters up and down by one standard deviation 
for the simulated samples, and assigning the difference in the measured mass as a systematic uncertainty.

\textbf{Systematic uncertainty in $\bm{p_T}$ resolution of muons}.
We estimate the uncertainty by changing the muon $p_T$ resolution~\cite{Abazov:2013xpp}
by $\pm1$  standard deviation  in the simulated samples 
and assign the difference in the measured mass as a systematic uncertainty.

\textbf{Jet identification}. Scale factors are used to correct the jet identification efficiency in MC events.
We estimate  the systematic uncertainty by changing these scale factors by $\pm1$ standard deviation. 

\textbf{Systematic uncertainty in jet resolution}.
The procedure of correction of jet energies for residual differences in energy resolution and energy scale in simulated events~\cite{Abazov:2013hda} applies additional smearing to the MC jets in
order to account for the differences in jet $p_T$ resolution
in data and MC.
To compute the systematic uncertainty on the jet 
resolution, the parameters for jet energy smearing
are changed by their uncertainties.

$\bm{b}$\textbf{-tagging efficiency}.
A difference in $b$-tagging modeling between data and simulation 
may case a systematic change in \mtop. To estimate this uncertainty, we change the $b$ tagging corrections up and down within their uncertainties using reweighting.

\subsection{Method}
\label{sec:Methodsyst}
\textbf{MC calibration}.
An estimate of the statistical uncertainties from the limited size of MC samples used in
the calibration procedure is obtained through the
statistical uncertainty of the calibration parameters.
To determine this contribution, we propagate the uncertainties on the calibration constants
$p_0$ and $p_1$ (Fig.~\ref{fig:calib}) to \mtop.

\textbf{Instrumental background}. To evaluate systematic uncertainty due to instrumental background, we change its contribution by $\pm 25\%$. The number of signal \ttbar\ events is recalculated by subtracting the instrumental background from the number of events in data, and ensemble studies are repeated to extract \mtop.

\textbf{Background contribution (or signal fraction)}.
To propagate the uncertainty associated with the background level, 
we change the number of background events according to its uncertainty, 
rerun the ensembles, and extract \mtop. 
In the ensembles, the number of \ttbar\ events is defined by the difference in the observed number of events in data and the expected number of background events.

\subsection{MC statistical uncertainty estimation}
We evaluated MC statistical uncertainties in the estimation of systematic uncertainties.
To obtain the MC statistical uncertainty in the \ttbar\ samples, we divide each sample into independent subsets.  
The dispersion of masses in these subsets is used to estimate the uncertainty.  The estimated MC statistical uncertainties for the signal modeling and jet and electron energy resolution  are $0.11 - 0.14$~GeV, for all other the typical uncertainty is around  $ 0.04$~GeV. In cases when the obtained estimate of MC statistical uncertainty is larger than the value of the systematic uncertainty, we take the MC statistical uncertainty as the systematic uncertainty.

\subsection{Summary of systematic uncertainties}
Table~\ref{tab:SummarySyst_AllwBkg} summarizes all contributions to the uncertainty on 
the \mtop\ measurement with the ME method.
Each source is corrected for the slope of the calibration from Fig.~\ref{fig:calib}(a).
The uncertainties are symmetrized in the same  way as in the \ljets\  measurement~\cite{Abazov:2014dpa,Abazov:2015spa}.
We use sign $\pm$ if the positive variation of the source of uncertainty corresponds to a positive variation 
of the measured mass, and  $\mp$ if it corresponds to a negative variation for two-sided uncertainties. We quote the uncertainties for one sided sources or the ones dominated by one-side component in Table~\ref{tab:SummarySyst_AllwBkg}, indicating the direction of \mtop change when using an alternative instead of the default model. 
 As all the entries in the total systematic uncertainty are independent,  the total systematic uncertainty on the top mass measurement is obtained by adding all the contributions in quadrature. 
 
\begin{table}
\begin{center}

\noindent
\begin{tabular}[t]{l|c}
\hline \hline
Source & Uncertainty (GeV)  \\ \hline
\textit{Signal and background modeling:} &  \\ 
Higher order corrections  &  $+0.16$ \\
ISR/FSR &  $ \pm 0.16$ \\
Hadronization and UE &  $+0.31$ \\
Color Reconnection  &  $+0.15$ \\
\textit{b}-jet modelling & $ +0.21$ \\ 
PDF uncertainty &  $ \pm 0.20$ \\
Heavy flavor &  $ \mp 0.06$ \\
$p_{T} (t\bar t) $ &  $ + 0.03$ \\
Multiple \ppbar\ interactions &   $ - 0.10$ \\

\hline
\textit{Detector modeling:} &  \\ 
Residual jet energy scale &  $ -0.20$  \\
Uncertainty on $k_{\rm JES}$ factor &  $\mp 0.46$ \\
Flavor dependent jet response &  $ \mp 0.30$ \\
Jet energy resolution &  $ \mp 0.15$   \\
Electron momentum scale & $\mp 0.10$ \\
Electron resolution &  $ \mp 0.16$ \\
Muon resolution & $\mp 0.10$ \\
\textit{b}-tagging efficiency &  $ \mp 0.28$ \\
Trigger & $\pm 0.06$ \\
Jet ID & $ +0.08$ \\
\hline
\textit{Method:}& \\ 
MC calibration  &  $ \pm 0.03$ \\
Instrumental background &  $ \pm 0.07$ \\
MC background  &  $ \pm 0.06$ \\

\hline\hline
Total systematic uncertainty  & $\pm 0.88$ \\
Total statistical uncertainty  & $\pm 1.61$ \\
\hline
Total uncertainty & $\pm 1.84
$ \\
\hline \hline
\end{tabular}


\caption{Systematic and statistical uncertainties for the measurement of $m_{t}$ in dilepton final states. The values are given for the combination of the $ee$, $e\mu$, and $\mu\mu$ channels.}           
\label{tab:SummarySyst_AllwBkg}
\end{center}
\end{table}

\section{Conclusion}
We have performed a measurement of the top quark mass in the dilepton channel $ t\bar{t} \to  W^+b\ W^-\bar{b} \to \ell^+\nu_\ell b\ \ell^- \bar{\nu_\ell} \bar{b}$\  using the matrix element technique  in \lumi\  of integrated luminosity collected by the D0 detector at the Fermilab Tevatron \ppbar\ Collider.
The result
$m_t = \mmeas \pm \mstat\ {\rm (stat)} \pm \msyst\ {\rm (syst)}$~\GeV,  corresponding to a relative precision of 1.0\%, 
 is consistent  with the values of the current Tevatron~\cite{Tevatron:2014cka} and world combinations~\cite{ATLAS:2014wva}.
\\


%

We thank the staffs at Fermilab and collaborating institutions,
and acknowledge support from the
Department of Energy and National Science Foundation (United States of America);
Alternative Energies and Atomic Energy Commission and
National Center for Scientific Research/National Institute of Nuclear and Particle Physics  (France);
Ministry of Education and Science of the Russian Federation, 
National Research Center ``Kurchatov Institute" of the Russian Federation, and 
Russian Foundation for Basic Research  (Russia);
National Council for the Development of Science and Technology and
Carlos Chagas Filho Foundation for the Support of Research in the State of Rio de Janeiro (Brazil);
Department of Atomic Energy and Department of Science and Technology (India);
Administrative Department of Science, Technology and Innovation (Colombia);
National Council of Science and Technology (Mexico);
National Research Foundation of Korea (Korea);
Foundation for Fundamental Research on Matter (The Netherlands);
Science and Technology Facilities Council and The Royal Society (United Kingdom);
Ministry of Education, Youth and Sports (Czech Republic);
Bundesministerium f\"{u}r Bildung und Forschung (Federal Ministry of Education and Research) and 
Deutsche Forschungsgemeinschaft (German Research Foundation) (Germany);
Science Foundation Ireland (Ireland);
Swedish Research Council (Sweden);
China Academy of Sciences and National Natural Science Foundation of China (China);
and
Ministry of Education and Science of Ukraine (Ukraine).

\bibliographystyle{apsrev_custom2}
\bibliography{References}


\end{document}